# NDSHA: ROBUST AND RELIABLE SEISMIC HAZARD ASSESSMENT

G. F. Panza[1]

## ABSTRACT

The Neo-Deterministic Seismic Hazard Assessment (NDSHA) method is found to reliably and realistically simulate the complete suite of earthquake ground motions that may impact civil populations as well as their heritage buildings. NDSHA's modeling technique is developed from comprehensive physical knowledge of: (i) the seismic source process; (ii) the propagation of earthquake waves; and (iii) their combined interactions with site effects. Thus, NDSHA effectively accounts for the *tensor* nature of earthquake ground motions: (a) formally described as the tensor product of the *earthquake source* functions and the *Green's Functions* of the transmitting (pathway) medium: and (b) more informally described as mathematical arrays of numbers or functions (indices) "that transform according to certain rules under a *change of coordinates.*"
Importantly, NDSHA therefore uses *all* available information about the *space distribution* of large magnitude earthquake phenomena, including: (a) Maximum Credible Earthquake (MCE) – which is based on seismic history and seismotectonics; and (b) geological and geophysical data. Thus it does not rely on *scalar* empirical ground motion *attenuation* models, as these are often both: (a) weakly constrained by available observations; and (b) fundamentally unable to account for the *tensor* nature of earthquake ground motion.
Standard NDSHA provides robust and safely conservative hazard estimates for engineering design and mitigation decision strategies; but without requiring (often faulty) *assumptions* about the *probabilistic risk analysis* model of earthquake occurrence. If specific applications may benefit from temporal information, including a gross estimate of the average occurrence time, the definition of the Gutenberg-Richter (GR) relation is performed according to the multi-scale seismicity model and occurrence rate is associated to each NDSHA modeled source.
Observations from recent destructive earthquakes in Italy, e.g. Emilia 2012, Central Italy 2009 and 2016-2017 seismic crises; and Nepal 2015 – have (a) confirmed the validity of NDSHA's approach and application; and (b) suggest that more widespread application of NDSHA will enhance earthquake safety and resilience of civil populations in all earthquake-prone regions, but especially in those *tectonically active* areas where the historic earthquake record is too short to have yet experienced (due to a relatively prolonged quiescence) the full range of *large, major* and *great earthquake* events that may potentially occur.

## Introduction

A new *Paradigm* is needed if Disaster Risk Mitigation is to succeed in fulfilling its very worthy goals! Mirroring the cautions and warnings of dozens of earlier papers (e.g. Molchan et al., 1997; Nekrasova et al., 2011; Panza et al., 2012; Bela, 2014), most recently Geller et al. (2016) and Mulargia et al. (2017) have concluded: (i) that everyone involved in seismic safety concerns should acknowledge the demonstrated *shortcomings* of PSHA (Probabilistic Seismic Hazard Analysis); (ii) that its use as a sacrosanct and unquestioningly-relied-upon *black box* for civil protection and public well-being must *cease*; and (iii) that most certainly a *new paradigm* is needed!
The Neo-deterministic Seismic Hazard Assessment methodology, NDSHA, described in detail by Panza et al. (2001; 2012), supplies a more *scientifically based* solution to the problem of more *reliably* characterizing earthquake hazard. Objective testing has never corroborated the validity of PSHA, which purports "to quantify the rate (or probability) of exceeding various ground-motion levels at a site (or a map of sites), given all possible earthquakes". The numerical/analytical approach to PSHA was first formalized by Cornell (1968) and Panza et al. (2014) have recently proven PSHA to be *unreliable!* Nevertheless, in spite of all the scientific and mathematical arguments and objections against it, PSHA (e.g.

[1]*Professor, Accademia Nazionale dei Lincei, Rome, Italy*
*Email – giulianofpanza@fastwebnet.it*

http://www.opensha.org/sites/opensha.org/files/PSHA_Primer_v2_0.pdf) has been widely used for almost 50 years by governments and industry when: (a) deciding safety criteria for nuclear power plants; (b) making official national earthquake hazard maps; (c) developing building code earthquake design standards; and (d) determining earthquake insurance rates.

As in any branch of science and physics, theories of earthquake occurrence should be *tested* and also *revised* in light of new data and experiences. In other words "theories *unsupported* by observations and experiments must be *corrected* or *rejected*, however intuitively appealing they might be." For example, many widely held beliefs with respect to earthquake occurrence (including timing and magnitude), such as Reid's 1906 elastic rebound theory as well as the *characteristic earthquake model*, unfortunately disagree with data (e.g. Molchan et al., 1997; Nekrasova et al., 2011; Kagan et al., 2012; and Geller et al., 2016). Therefore incorporation of such invalid "implicit assumptions" in models, which then make *probabilistic* statements about future near term seismicity, as is the case for PSHA, helps make these models even more untestable: (i) on either a local scale; (ii) on a regional scale; and (iii) within a realistic time scale (e.g., Beauval et al., 2008; Panza et al., 2014)

PSHA, because it has too often delivered not only erroneous but also too deadly results, has been extensively debated over many years; and a sample of contributions is contained in the PAGEOPH Topical Volume 168 (2011) and references therein. NDSHA, over now some two decades, provides both reliable and effective earthquake hazard assessment tools for understanding and mitigating earthquake risk. Unlike PSHA, NDSHA *is falsifiable* – however it has been, to the contrary, very effectively validated by all earthquake events which have occurred in regions where NDSHA hazard maps were already available at the time of the quake: (i) Emilia-Romagna (Peresan and Panza, 2012); (ii) Central Italy (Fasan et al., 2016); and (iii) Nepal (Parvez et al., 2017).

It should be widely taken to heart, I believe, that the *continued* practice of PSHA for determining earthquake resistant design standards for civil protection, mitigation of heritage and existing buildings, and community economic well-being and resilience . . . is in a *state-of-crisis!* And alternative methods, which are already available and ready-to-use, like NDSHA, should be applied worldwide. The results will then be twofold: (1) to extensively test these alternative methods; and (2) to prove that they globally actually perform more reliably (safely) than PSHA . . . R.I.P! In fact, even knowing MCE, to obtain the maximum acceleration by means of attenuation relations is a PSHA daydream that continues to lead to social nightmare.

## Methodology

The procedure for the neo-deterministic seismic hazard assessment (NDSHA) (Panza et al., 2001; 2012) is based on the computation of realistic synthetic seismograms (ground motion scenarios). In NDSHA, seismic hazard is defined as the envelope of the values of earthquake ground motion parameters considering a wide set of scenario events, including maximum credible earthquakes (MCE), calculated by means of physically rooted models and using the available physical knowledge on earthquake sources and wave propagation processes. NDSHA does not rely on empirical attenuation models, often weakly constrained by the available observations and unable to account for the tensor nature of earthquake ground motion, that, as it is well-known, is formally described as the tensor product of the earthquake source tensor with the Green's function of the medium (Aki and Richards, 2002).

NDSHA (see Figure 1) belongs to the class of methods that employ numerical modeling codes based upon physical description of earthquake rupture process and seismic wave propagation to reliably predict the ground motion parameters from the potential seismic sources (see Panza et al., 2001; 2012; Magrin et al., 2016; Panza et al., 2013 and the references therein). NDSHA stands for scenario-based methods for seismic hazard analysis: starting from the available knowledge about Earth's structure through which seismic waves propagate, seismic sources and seismicity of the study area, it is possible to realistically compute the synthetic seismograms from which one can quantify peak values of acceleration (PGA), velocity (PGV) and displacement (PGD) or any other ground motion parameter relevant to seismic engineering, e.g. design ground acceleration (DGA) computed consistently with the shape of any preferred design spectrum. The DGA is the acceleration anchoring the elastic response spectrum at period $T = 0$ s. This quantity is comparable to PGA, since an infinitely rigid structure (i.e., a structure having a natural period of 0 s) moves exactly like the ground (i.e., the maximum acceleration of the structure is the same as that of the ground, which is the PGA). Moreover, DGA is practically equivalent to effective peak acceleration (EPA), which is defined as the average of the maximum ordinates of elastic acceleration response spectra within the period range from 0.1 to 0.5 s, divided by a standard factor of 2.5, for 5% damping (Panza et al., 2004).

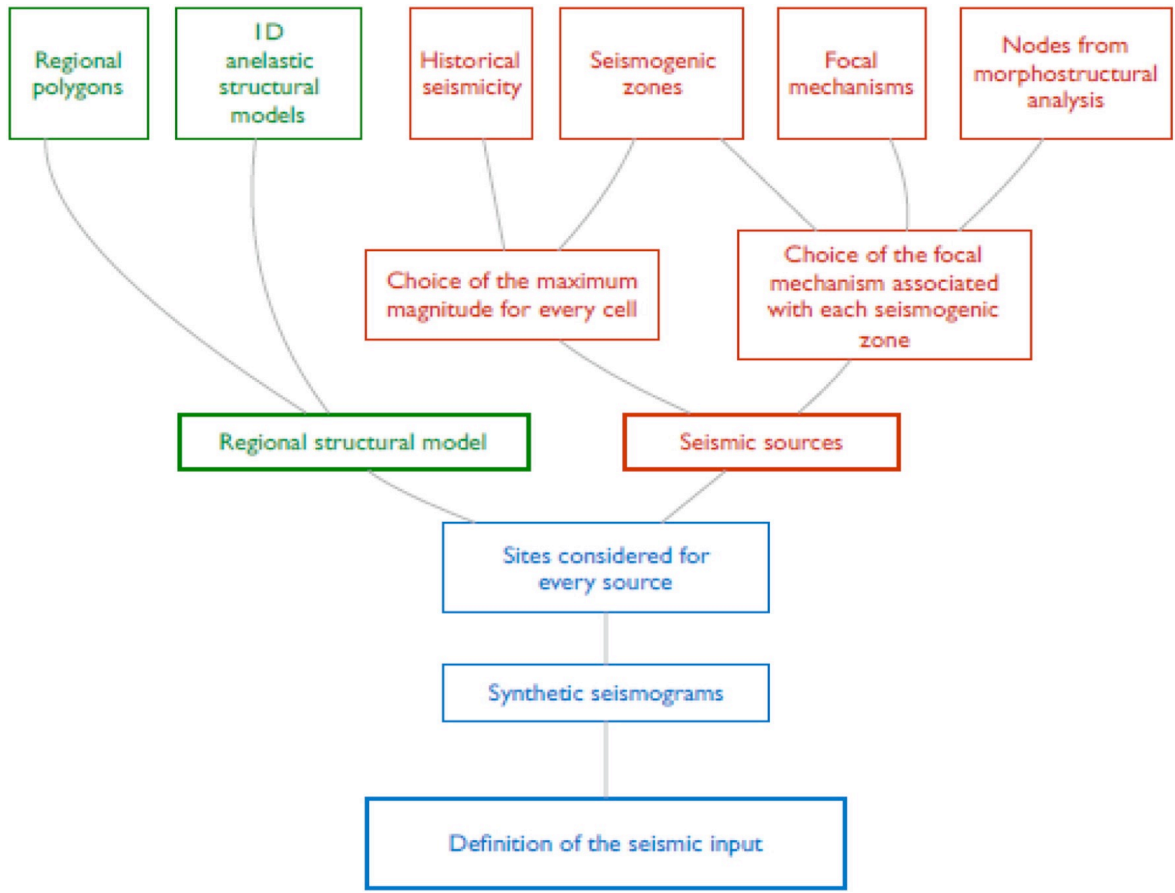

**Figure 1.** Flow chart of NDSHA at regional scale.

As a rule, a regular grid (usually 0.2° × 0.2°) is placed over the study region. The earthquake sources are centered in the grid cells that fall within the adopted seismogenic zones, while the computation sites are placed at the nodes of a grid that is staggered by 0.1° with respect to the sources' grid. A smoothing procedure for the definition of earthquakes location and magnitude, M, is then applied to partly account for spatial uncertainty, catalogue incompleteness and for earthquake source extension. After smoothing only the cells (earthquake sources) located within the seismogenic zones or within a seismogenic node are retained. A double-couple point source is placed at the center of each cell, with a representative focal mechanism, which is consistent with the known present-day dominant tectonic regime of the corresponding seismogenic zone. To define the magnitude of each source, the NDSHA procedure makes use of information about the space distribution of large-magnitude earthquakes (M > 5), which can be defined from historical, instrumental and geological observations. The source depth is taken into consideration as a function of magnitude, in agreement with literature (e.g. Caputo et al., 1973; Molchan et al., 1997; Doglioni et al., 2015). A complete description of the NDSHA methodology can be found in Panza et al. (2001) and its updates and validations in Panza et al. (2012), Fasan et al. (2016), Magrin et al. (2016), Fasan (2017) and Hassan et al. (2017).

In addition, NDSHA permits, if really necessary, to account for earthquake occurrence rate (Peresan et al., 2013 and references therein; Peresan et al., 2014; Magrin et al., 2017). Peresan et al. (2013) have performed the characterization of the frequency-magnitude relation for earthquake activity in Italy according to the multi-scale seismicity model (Molchan et al., 1997; Kronrod, 2011), so that a robust estimated occurrence is associated to each of the modeled sources. The occurrence assigned to the source is thus associated to the pertinent synthetic seismograms, coherently with the physical nature of the problem. Accordingly two separate maps are obtained: one for the ground shaking, one for the corresponding occurrence. In fact when considering two sites prone to earthquakes with the same magnitude M, given that all the remaining conditions are the same, the parameters for seismic design must be equal at the two sites, since the magnitude we have to defend against is the same, independently from the sporadic occurrence of the earthquake. The flow chart that describes the NDSHA procedure for regional scale analysis is shown in Fig. 1. The physics-based ground motion modeling is limited up to a frequency of 10 Hz because the estimate of ground motion at higher frequencies requires the knowledge of source heterogeneity, physical properties of the rock/soil and the attenuation parameters with a resolution realistically not attainable. This is well in agreement with Aki's (2003) conclusion: results about the source-controlled $f_{max}$, non-linear soil response and the studies of seismic attenuation from borehole data indicate that there is no need to consider frequencies higher than about 10 Hz in strong motion seismology. In fact, the quality of the results obtained by physics based ground motion modeling depends on the quality of the input data. The NDSHA approach allows for sensitivity analyses to evidence and address the uncertainties using different input data and varying level of knowledge about seismic sources and structural models. The proper presentation and evaluation of uncertainties, associated with the ground motion computation, will help the potential users to determine how much confidence to place on the estimated seismic hazard map. The strength of the source is determined as the maximum between a lower bound and the magnitude defined by the smoothing procedure. The lower bound for magnitude inside the seismogenic zones is 5, that is conventionally (D'Amico et al., 1999) taken as the lower bound for the magnitude of damaging earthquakes. The lower bound of magnitude inside the seismogenic nodes is the magnitude threshold identified for that node by the morphostructural analysis (Gelfand et al., 1972). The orientation of the double-couple point source is the one representative of the parent seismogenic zone or seismogenic node. Hypocentral depth, in fairly good agreement with existing literature, is taken as a function of magnitude (10 km for $M < 7$, 15 km for $7 \leq M < 8$ and 25 km for $M \geq 8$). More and updated details about the procedure can be seen in Parvez et al. (2017).

## Validations by facts

NDSHA is falsifiable. The detailed review of the traditional PSHA method (besides just implementation errors) revealed that the method is not adequate to describe the physical process of earthquake occurrence because of the assumption of a memory less stochastic process – Poisson process. It is obvious that strain and stress renewal needs time and therefore the process of rebuilding the conditions for the next earthquake is time dependent. Furthermore, the location of earthquakes even at the same fault is changing with time as well as its mechanical properties, in particular after each event. Each big earthquake is modifying the boundary conditions for the next one. This means that a mathematical probabilistic model has to be at least bivariate. This is outside of the scope of human knowledge due to lack of data and the shortness of human observation time in comparison with geological ages.

As any physical model NDSHA suffers of uncertainties and limitations due to the uncertainty intrinsic in the basic data, chiefly earthquake catalogues, and lack of satisfactory theories about earthquake source. For this reason hazard values supplied by NDSHA are given as ranges over areas whose values are consistent with the information content of the basic data. Typical values are grid mesh about 25-50 km and hazard values in discrete ranges in geometrical progression close to 2 (Figure 2).

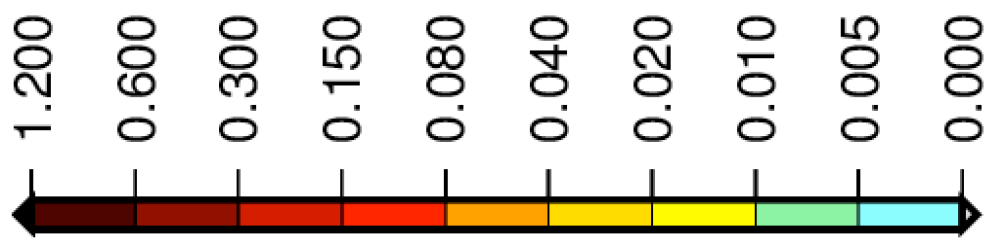

**Figure 2.** Typical discrete ranges of hazard values (units of g), in geometrical progression close to 2, consistent with the real resolving power of the worldwide available data (e.g. Cancani, 1904; Lliboutry, 2000).

NDSHA is solving most of the problems posed by an adequate description of the physical process of earthquake occurrence. It takes the largest event physically possible, usually termed maximum credible earthquake (MCE), whose magnitude can be tentatively and until proven otherwise set equal to the maximum observed magnitude plus some multiple of the standard deviation ($s_M$), and the shortest known distance from fault to site and compile and envelope spectra (S) from all known pertinent sources. In areas without information on faults or sparse data, historical and morphostructural (Gelfand et al., 1972; for a recent application Gorshkov et al., 2012) data are used; these same data can be used to enrich existing parametric earthquake catalogues (Zuccolo et al., 2011; Parvez et al., 2017). To encompass uncertainty and in dependence on the safety importance of structures the resulting broad band spectrum S can be increased by a quantity obtained multiplying its standard deviation ($s_S$) by a safety factor (usually variable between 1.4 and 1.6) augmented by one unit, i.e. about 2,4-2,6($s_S$). This choice is consistent with Chebyshev's theorem: for a very wide class of probability distributions, no more than a certain fraction of values can be more than a certain distance from the mean. Specifically, no more than $1/k^2$ of the distribution's values can be more than $k$ standard deviations away from the mean (or equivalently, at least $1-1/k^2$ of the distribution's values are within $k$ standard deviations of the mean). If $k=2$ at least 75% of the values fall within 2$s$ and if $k=3$ at least 89% of the values fall within 3$s$.

NDSHA aims to supply an envelope value, in other words a value that should not be exceeded, therefore it is immediately falsifiable/verifiable: if an earthquake occurs with a magnitude larger than that indicated by NDSHA it is necessary to measure the variance. If the variance turns out to be larger than the multiple of standard deviation ($s_M$) used to define MCE, for instance (Dominique and Andre, 2000) equal to the maximum observed magnitude plus $2s_M = 0.5$ (a value which is representative of twice the standard deviation of magnitude determination at global scale), then maps are immediately falsified; similarly if the peak values (e.g. PGA) recorded at the bedrock at the occurrence of an earthquake after the compilation of NDSHA maps exceed, within error limits, those given in the same maps. On the contrary, the falsification of PSHA maps, that may be eventually possible only considering very large seismogenic zones and time intervals (Molchan et al., 1997; Panza et al., 2014), does not supply any useful information on account of the ambiguity of the interpretation of the violation: "Bad assumptions or bad luck?" (Stein et al., 2012).

## Study cases

To better describe from a practical point of view limits and possibilities of NDSHA, let us now consider a few study cases.

### Northern Italy - The Emilia earthquake crisis in 2012

Currently, the PSHA map from Gruppo di Lavoro, Redazione della mappa di pericolosità sismica, rapporto conclusivo, 2004, (http://zonesismiche.mi.ingv.it/mappa_ps_apr04/italia.html) is the official reference seismic hazard map for Italy, and shows bedrock PGA values that have a 10% probability of being exceeded in 50 years (i.e. once in 475 years). The Emilia 20 May 2012 M=5.9 and 29 May 2012 M=5.8 earthquakes occurred in a zone defined at low seismic hazard by the code based on PSHA: PGA map ("return period" 475 yr) < 0.175g; observed PGA > 0.25g. The NDSHA map published in 2001 (Panza et al., 2001), which expresses shaking in terms of design ground acceleration, DGA, equivalent to peak ground acceleration, PGA, (see Zuccolo et al. 2011), predicted values in the range 0.20–0.35g, in good agreement with the observed motion that exceeded 0.25g. Seismic hazard maps seek to predict the shaking that would actually occur, therefore what occurred in Northern Italy supplies a strong motivation for the use NDSHA or similar deterministic approaches, also with the aim to minimize the necessity to revise hazard maps with time. In this view, public buildings and other critical structures should be designed to resist future earthquakes. Contrary to what implicitly suggested by PSHA, when an earthquake with a given magnitude occurs, it causes a specific ground shaking that certainly does not depend on how sporadic the event is (rare or not). Hence ground motion parameters for seismic design should be independent of how sporadic (infrequent) an earthquake is, as it is done with NDSHA (Peresan and Panza, 2012 and references therein).

### Central Italy

#### *The Aquila 2009 event*

The 6 April 2009 M=6.3 earthquake occurred in a zone defined at high seismic hazard, but the observed acceleration values exceeded those predicted by the code based on PSHA: PGA map (475 yr) 0.250-0.275g; observed PGA > 0.35g. The NDSHA map predicts values in the range 0.3-0.6g and this implies that future events may cause peak ground motion values exceeding those recorded in 2009. As far as I know such obvious caution is not explicitly, duly considered in the ongoing reconstruction.

*The earthquake crisis started in 2016*

The 24 August M=6.0 and 30 October M=6.5 earthquakes occurred in a zone defined as high seismic hazard, but the observed acceleration values exceeded those predicted by the code standards, based on PSHA: PGA map (475 yr) 0.250-0.275g; observed PGA > 0.4g (a value that is larger than the one recorded at l'Aquila). Alternatively, the NDSHA map predicts values in the range 0.3-0.6g. Following the earthquake crisis, starting August 24, 2016 supplies a dramatic example of a familiar English proverb "you get what you pay for", and maybe even "a little bit of knowledge is a dangerous thing;" as described in the following section (*The lesson*).

*The lesson*

After these recent events many civil engineers, designers and practitioners complained about the fact that the accelerations given in the code standard hazard maps based on PSHA are *distorted downwards*. The due revision of these maps therefore due (to coincide with real earthquake experiences), however, encounters strong and blunt resistance, particularly regarding the *methodological* approach of PSHA. However, more than just these methodological aspects, it is also necessary to seriously consider the contrast between the very real benefits of risk reduction, as against the building costs increment - if more *reliable* and *robust risk coefficients* are considered (consistent with the magnitude size of *possible* events, like these recent ones). Beyond the untold human toll, *a posteriori* retrofitting costs about 30 times more than the upgrading of earthquake resistant design standards at the time of *new* construction (here to the more realistic and therefore more stringent earthquake resistance measures as identified by NDSHA.

Thus it looks very appropriate to remember the wisdom of the ancient Greek teachings and more modern everyday proverbs: "adequate prevention is better than cure" – as Hippocrates said about 2500 years ago; and the proverb "you get what you pay for". A dramatic example is provided by the city of Norcia.

Norcia had been retrofitted after the Umbria-Marche earthquake crisis started September 26, 1997. All reconstruction works used as a benchmark the PSHA map (475 yr) on which the seismic code was based. Those maps proved totally misrepresentative and erroneous at the occurrence of the 30 October 2016 M=6.5 earthquake, where in Norcia the earthquake ground motion was much larger than what had been predicted by PSHA. The resulting damage was large, corresponding to $I_{MCS}$=IX (http://www.6aprile.it/wp-content/uploads/2016/12/QUEST_rapporto_15nov.pdf. - reports $I_{MCS}$=VII-IX, but it should be kept in mind that any intensity scale is discrete with unit incremental step). On the NDSHA map, the hazard value indicated is slightly above the experienced ground motion generated by the 30 October 2016 earthquake. In all likelihood, if the reconstruction and retrofitting that followed the 1997 Umbria-Marche earthquakes would have been undertaken in due account of the NDSHA estimates, the damage would have been much less (if not negligible) with respect to that *actually observed* after the (30/10/2016) event.

To have followed PSHA designated design strength and detailing requirements for new buildings, while neglecting that the Italian seismic code further provides: "*L'uso di accelerogrammi generati mediante simulazione del meccanismo di sorgente e della propagazione è ammesso a condizione che siano adeguatamente giustificate le ipotesi relative alle caratteristiche sismogenetiche della sorgente e del mezzo di propagazione*" (NTC 2008 chapter 3.2.3.6) ["The use of accelerograms generated simulating source mechanism and wave propagation is allowed provided the hypotheses about the seismogenic characteristics of the source and the properties along the pathway are duly justified."] – certainly allowed some (marginal) cost saving during the reconstruction and retrofitting following the 1997 events, when compared to the higher earthquake resistance requirements indicated under NDSHA. Nonetheless this apparent "saving" has been unrealized and ultimately *frustrated* by the October 2016 earthquake – and now it is necessary to consider in the reconstruction and retrofitting the NDSHA values, which were unwisely *ignored* after the 1997 earthquakes.

Lastly to consider (but not least), *before* the occurrence of 30 October 2016 the M=6.5 event, when Norcia was almost completely destroyed:
  (a) Fasan et al. (2016) did show that the spectral accelerations for the 30 October 2016 M=6.5 event, with magnitude close to the maximum ever historically observed in the area, are in very good agreement with what had earlier been predicted, based on NDSHA ground motion simulations;
  (b) Panza and Peresan (2016) issued the warning that the 24 August 2016 M=6.0 earthquake did not

necessarily generate the largest possible ground motion in the area: since the area had been previously hit by the 14 January 1703 M=6.9, Valnerina earthquake. They further warned that, in the ensuing reconstruction and retrofitting activity, engineers should take into account as well that, in the future, seismic source and local soil effects may lead to ground motion values exceeding the NDSHA value of 0.6g (predicted at the bedrock).

Therefore many now believe that it is well validated and justified to claim that NDSHA is a reliable and ready alternative to the presently widespread use of PSHA, particularly since its use has been widely proven in the professional journals and publications to be a totally unjustified and unreliable procedure (e.g. Klügel, 2008; PAGEOPH Topical Volume 168, 2011; Mulargia et al., 2017; Fasan, 2017).

**India**

The Neo-deterministic Seismic Hazard Assessment, expressed in terms of maximum displacement (Dmax), maximum velocity (Vmax) and design ground acceleration (DGA), has been extracted from the synthetic seismograms and mapped on a regular grid of 0.2° × 0.2° over the entire country (Parvez et al., 2017). The highest seismic hazard, expressed in terms of DGA (in the range of 0.6-1.2g), is mainly distributed: (i) in western Himalayas and Central Himalayas along the epicentral zone of the Bihar Nepal 2015 earthquake; (ii) part of NE India and (iii) in the Gujarat (Kachchh region). A similar pattern has been found in peak velocities and peak displacements in the same areas. For the same event, using the conversion acceleration to EMS intensity (Llibroutry, 2000), the NDSHA results have been compared with the maximum observed intensities reported in EMS scale by Martin and Szeliga (2010): where observations are available, the modeled intensities are rarely exceeded by the maximum observed intensities.

**Conclusions**

NDSHA has, for over now *two decades*, provided both *reliable* and *effective* earthquake hazard assessment tools for understanding, communicating and mitigating earthquake risk (Panza et al., 2001). And the NDSHA procedure for the development of seismic hazard maps at the regional scale is described in some detail at http://www.xeris.it/Hazard/index.html. Moreover, NDSHA seismic hazard assessment has been well validated by all events occurring in regions where NDSHA maps were available at the time of the quake; including these observations from four recent destructive earthquakes: M 6 Emilia, Italy 2012; M 6.3 L'Aquila, Italy 2009; M 5.5-6.6 Central Italy 2016-17 seismic crises; and M 7.8 Nepal 2015. This good performance suggests that the wider adoption of NDSHA (especially in tectonically active areas – *but* with relatively prolonged quiescence, i.e. where only few major events have occurred in historical time) can better prepare civil societies for the entire suite of potential earthquakes that can . . . and will occur!  Better to retire and then *bury* PSHA, which is more *concept* than it is a tested pathway to seismic safety, R.I.P . . . than to experience *more* earthquake disasters and catastrophes, where erroneous hazard maps depicted only "low hazard," but the active tectonic regions acted otherwise!

PSHA*, unlike NDSHA*, has: (1) never been validated by "objective testing;" (2) actually been proven *unreliable* (Panza et al., 2014) as a forecasting method on the rates (but claimed *probabilities*) of earthquake occurrence; and (3) staked its hype and dominance on *assumptions* that both *earthquake resistant design standards* and *societal earthquake preparedness and planning* should be based on "engineering seismic risk analysis" models – models which incorporate assumptions, really *fabulations* (or "magical realisms") now known to conflict with what we have learned *scientifically* regarding *earthquake geology* and *earthquake physics* over the *same* (almost 50-yr) time frame of PSHA's: (i) initial hype; (ii) acceptance; and (iii) eventual 40-yr rise to dominance).PSHA, because it has too often delivered not only erroneous but also too deadly results (Wyss et al., 2012; Panza et al., 2014; Bela, 2014), has been extensively debated over many years; a sample of contributions is contained in the PAGEOPH Topical Volume 168 (2011) and references therein.  In evidence against PSHA:  too many damaging and deadly earthquakes (like the 1988 M 6.8 Spitak, Armenia; the 2011 M 9 Tohoku, Japan Megathrust; and the 2012 M 6 Emilia, Italy events) have all occurred in regions rated to be "low-risk" by PSHA derived seismic hazard maps (e.g. Peresan and Panza, 2012; Mulargia et al., 2017.)

It should therefore be widely *taken-to-heart*, I believe, that the *continued* practice of PSHA for determining earthquake resistant design standards for civil protection, mitigation of heritage and existing buildings, and community economic well-being and resilience . . . is in a *state-of-crisis!*  And alternative methods, which are already available and ready-to-use, like NDSHA, should be applied worldwide.  The results will then be twofold: (1) not only to extensively *test* these alternative methods; but (2) to *prove* that they globally actually perform more reliably and safely than PSHA . . . R.I.P!

## Acknowledgments

The development and application of NDSHA would have been impossible without the committed engagement of many Colleagues, Post-doctoral Researchers, Graduate Students and Ph.D. Students, worldwide. To mention all of them would generate an almost endless list of names and countries, but scientifically not much informative. Much more information about their fundamental and multidisciplinary contributions can be obtained consulting with the due attention the reference's list of the many papers I co-authored with them. This paper offers a unique opportunity for me to express my heartfelt thanks to all of them. A special appreciation goes to Antonella Peresan and James Bela for the critical and very constructive reviewing of the original manuscript that led to this significantly improved text.